\documentclass[aps,prl,showpacs,raggedbottom, 
amssymb,twocolumn,groupedaddress]{revtex4-1}
\usepackage{graphicx}
\usepackage{amsmath}
\usepackage{amsfonts}
\usepackage{amssymb}
\usepackage{epsfig,libertine}
\usepackage[libertine]{newtxmath}
\usepackage{color}
\usepackage{dsfont, hyperref, xcolor}
\usepackage{comment}
\usepackage{enumerate}

\newcommand{\abs}[1]{\left\vert#1\right\vert}

\newcommand{\bra}[1]{\left\langle#1\right\vert}
\newcommand{\ket}[1]{\left\vert#1\right\rangle}

\begin{document}

\title{Anomalous universal adiabatic dynamics: the case of the Fredkin model}

\author{Gianluca~Francica, Luca Dell'Anna}
\address{Dipartimento di Fisica e Astronomia ``G. Galilei'', Universit\`{a} degli Studi di Padova, via Marzolo 8, 35131 Padova, Italy}

\date{\today}

\begin{abstract}
When a system is driven across a second-order quantum phase transition, 
the number of defects which are produced scales with the speed of the variation of the tuning parameter according to a universal law described by the Kibble-Zurek mechanism. We study a possible breakdown of this prediction proving that the number of defects can exhibit another universal scaling law which is still related only to the critical exponents $z$ and $\nu$, but differs from the Kibble-Zurek result. Finally we provide an example, the deformed Fredkin spin chain, where this violation of the standard adiabatic dynamics can occur.
\end{abstract}

\maketitle

{\it Introduction.} Near a quantum phase transition the behavior of some physical quantities is generally described only in terms of critical exponents~\cite{sachdevbook}. In particular, by driving the system across a second-order quantum phase transition by slowly changing a parameter, the density of defects that come out scales algebraically with the speed of the variation of the tuning parameter~\cite{zurek05,dziarmaga05,polkovnikov05}. In detail, the exponent of the scaling law depends only on the critical exponents $z$ and $\nu$, as predicted by the Kibble-Zurek mechanism~\cite{kibble76,zurek85}.
This phenomenon has been confirmed theoretically by using different approaches and observed experimentally in different physical systems (see, e.g., Refs.~\cite{polkovnikov11,delcampo14} and references therein).\\
Here, we prove that there are systems undergoing a universal dynamics different from the one described in Ref.~\cite{polkovnikov05}. In detail, we show that the exponent of the density of defects depends only on the critical exponents $z$ and $\nu$, but it is different from that predicted by the Kibble-Zurek mechanism. At the quantum phase transition, approaching the thermodynamic limit, these systems might have a finite size scaling which depends on the excited states. In this case and if this dependence is smooth, we have still an universal adiabatic dynamics, different from the Kibble-Zurek prescription.  We show that our theory applies to the deformed Fredkin spin chain~\cite{salberger17}. 
{\color{black}The Fredkin model, in principle, can be realized experimentally by cold atoms in optical lattices, and a three-spin interaction can be designed as in Ref. \cite{pachos}}.
The system is driven through a deformation parameter $g$ across a quantum critical point described by the Fredkin spin chain~\cite{dellanna16,salberger16}, which, however does not have a conformal invariance, being the dynamical exponent $z\geq 2$.

{\it Universal adiabatic dynamics.}  In order to derive our main result, we consider a one dimensional many-body system of size $N$ described by a Hamiltonian $H(\lambda)$ which shows a second-order quantum phase transition at $\lambda=0$, with critical exponents $z$ and $\nu$. If the parameter $\lambda$ changes linearly in time, like $\lambda(t) = t/\tau$, the time-evolved state can be expressed as a linear combination of the eigenstates $\ket{E_l(\lambda)}$ with eigenvalues $E_l(\lambda)$ (in a non-decreasing order), as $\ket{\psi(t)}= \sum_l c_l(t) e^{-i\int_{t_{in}}^t E_l(\lambda(t')) dt'} \ket{E_l(\lambda(t))}$. As shown in Ref.~\cite{polkovnikov05}, if the initial state is the ground-state $\ket{E_0(\lambda(t_{in}))}$, the relative number of adiabatic excitations $n_{ex} = \sum_{l>0} \abs{c_l(t)}^2$ can be expressed as
\begin{equation}\label{eq. 0}
n_{ex}=\sum_l \abs{ \int_{-\infty}^\infty d\lambda \bra{E_l(\lambda)}\partial_\lambda \ket{E_0(\lambda)} e^{i\tau \int_{-\infty}^\lambda d\lambda' (E_l(\lambda')-E_0(\lambda'))}}^2\,.
\end{equation}
%
Let us introduce the momenta $k_l$, 
such that both the scaling relations
\begin{eqnarray}\label{eq. sca e}
E_l(\lambda)-E_0(\lambda) &=& \lambda^{z\nu} \tilde F(\lambda^{z \nu}/k_l^z)\,,\\
\label{eq. sca} \bra{E_l(\lambda)} \partial_\lambda \ket{E_0(\lambda)} & =& \frac{\lambda^{z\nu-1}}{k_l^z} \tilde G\left(\frac{\lambda^{z\nu}}{k_l^z}\right)
\end{eqnarray}
hold, and if $k_l$ are homogeneously distributed around zero, $k_l\sim 0$, with proper universal scaling functions $\tilde F$ and $\tilde G$, 
we obtain the Kibble-Zurek scaling law, i.e., $n_{ex} \sim \tau^{-\gamma_0}$ as $\tau\to\infty$, with $\gamma_0 = \nu/(z\nu+1)$ \cite{polkovnikov05}. 
If, instead, the distribution of $k_l$ is $\rho(k)\sim k^\beta$ for $k\sim 0$, it is easy to see that $n_{ex} \sim \tau^{-(1+\beta)\gamma_0}$. \\
Let us now consider the case where Eq.~\eqref{eq. sca} is modified. Our aim is to understand how the number of excitations $n_{ex}$ scales with $\tau$ in this case. Due to Eq.~\eqref{eq. sca e}, we have that
\begin{equation}\label{eq. 1}
n_{ex} = \sum_l \abs{ \int d\lambda \bra{E_l}\partial_\lambda \ket{E_0} e^{i\tau \int^\lambda d\lambda \lambda^{z\nu}\tilde F(\lambda^{z \nu}/k_l^z)}}^2\,.
\end{equation}
Let us suppose that the main contribution to the integral comes only from small $k_l$, as it is usually the case, and assume that $|\bra{E_l}\partial_\lambda \ket{E_0}|$ has a global maximum at $\lambda_c(N)\sim N^{-1/\nu}$, which scales as $k_l^{-a_l}$ for $k_l\sim 0$. 
For $N\to \infty$, we expect that $\bra{E_l}\partial_\lambda \ket{E_0}$ is well approximated by a Gaussian in a neighborhood of $\lambda_c(N)$, i.e., $\bra{E_l}\partial_\lambda \ket{E_0}\sim k_l^{-a_l} e^{-(\lambda-\lambda_c(N))^2/(2\sigma_\lambda^2)}$.
We make the following finite size scaling ansatz \cite{domb83}
\begin{equation}
\bra{E_l}\partial_\lambda \ket{E_0} = N^{a_l} g_l(N^{1/\nu}(\lambda-\lambda_c(N)))\,.
\end{equation}
If this formula holds, then we get $\sigma_\lambda \sim k_l^{1/\nu}$. For instance, if the scaling law in Eq.~\eqref{eq. sca} is fulfilled we get a homogeneous critical exponent $a_l$, which is equal to $a_l=1/\nu$.
For $\tilde F(x) \sim 1/x$, considering the integral (up to a global phase)
\begin{equation}
\int d\lambda k_l^{-a_l} e^{-(\lambda-\lambda_c(N))^2/(2\sigma_\lambda^2) + i \tau k_l^{z}\lambda} \sim k_l^{1/\nu-a_l} e^{-c \tau^2 k_l^{2z+2/\nu}/2}\,,
\end{equation}
from Eq.~\eqref{eq. 1}, we get
\begin{equation}
n_{ex} \sim \sum_l A_l k_l^{2/\nu-2a_l} e^{-c\tau^2 k_l^{2z+2/\nu}}\,.
\end{equation}
Defining $a(k_l)=a_l$, we get, in the thermodynamic limit, 
\begin{equation}\label{eq. 2}
n_{ex} \sim \int \rho(k)  k^{2/\nu-2a(k)} e^{-c\tau^2 k^{2z+2/\nu}}dk\,.
\end{equation}
We note that $c$ defines the time scale, and we can consider $c=1$, without loss of generality, in order to calculate the limit $\tau\to\infty$. We consider $a(k)$ smooth such that $a(k)\sim a(0)+u k$ as $k\sim 0$, and we assume that the scaling formula of Eq.~\eqref{eq. sca} holds for the first excited state $l=1$, so that $a(0)=1/\nu$. If $u=0$, the integral of Eq.~\eqref{eq. 2} can be easily calculated, and, for $\beta=0$, we recover the Kibble-Zurek scaling law $n_{ex}\sim \tau^{-\gamma_0}$. Let us consider the case $u\neq 0$.
By performing the substitution $k=(\eta/\tau)^{\gamma_0}$, and extending the integral over $\eta$ from zero to infinity, for $\beta=0$ and $c=1$, from Eq.~\eqref{eq. 2} we get
\begin{equation}\label{eq. 3}
n_{ex} \sim  \int_0^\infty d\eta (\eta/\tau)^{\gamma_0} \eta^{-1} e^{f(\eta)}\,,
\end{equation}
where $f(\eta)=-2 u \gamma_0 (\eta/\tau)^{\gamma_0} \ln (\eta/\tau)-\eta^2$.
In order to calculate the asymptotic formula for $\tau\to \infty$, we use the Laplace method. For $u>0$, the function $f(\eta)$ has a global maximum at $\eta_0=x_0 \tau$, where
\begin{equation}
x_0 = \left( \frac{u \gamma_0^2}{(2-\gamma_0)\tau^2} W\left((2-\gamma_0)e^{1-2/\gamma_0} \tau^2/(\gamma_0^2 u)\right) \right)^{1/(2-\gamma_0)} ,
\end{equation}
where $W(x)$ is the Lambert W function defined such that $W(x) e^{W(x)}=x$. Then, we get
\begin{equation}
\label{nexu}
n_{ex} \sim \frac{1}{\sqrt{-f''(\eta_0)}} (\eta_0/\tau)^{\gamma_0} \eta_0^{-1} e^{f(\eta_0)}\,,
\end{equation}
where $f(\eta_0)= 2 u x_0^{\gamma_0} + (2/\gamma_0-1)\tau^2 x_0^2$, and $f''(\eta_0)= -2 u \gamma_0^2\tau^{-2} x_0^{\gamma_0-2}+2(\gamma_0-2)$.
For extremely slow driving, $\tau \to \infty$, we get $n_{ex} \sim \tau^{-\gamma}(\ln \tau)^{(\gamma-1)/2}$, where 
\begin{equation}\label{eq. gamma limit}
\gamma = - \lim_{\tau\to \infty} \frac{\tau \partial_\tau n_{ex}}{n_{ex}} = \frac{\gamma_0}{2-\gamma_0}\,,
\end{equation}
which is again a universal critical exponent depending only on $z$ and $\nu$.
\begin{figure}
[h!]
\centering
\includegraphics[width=0.75\columnwidth]{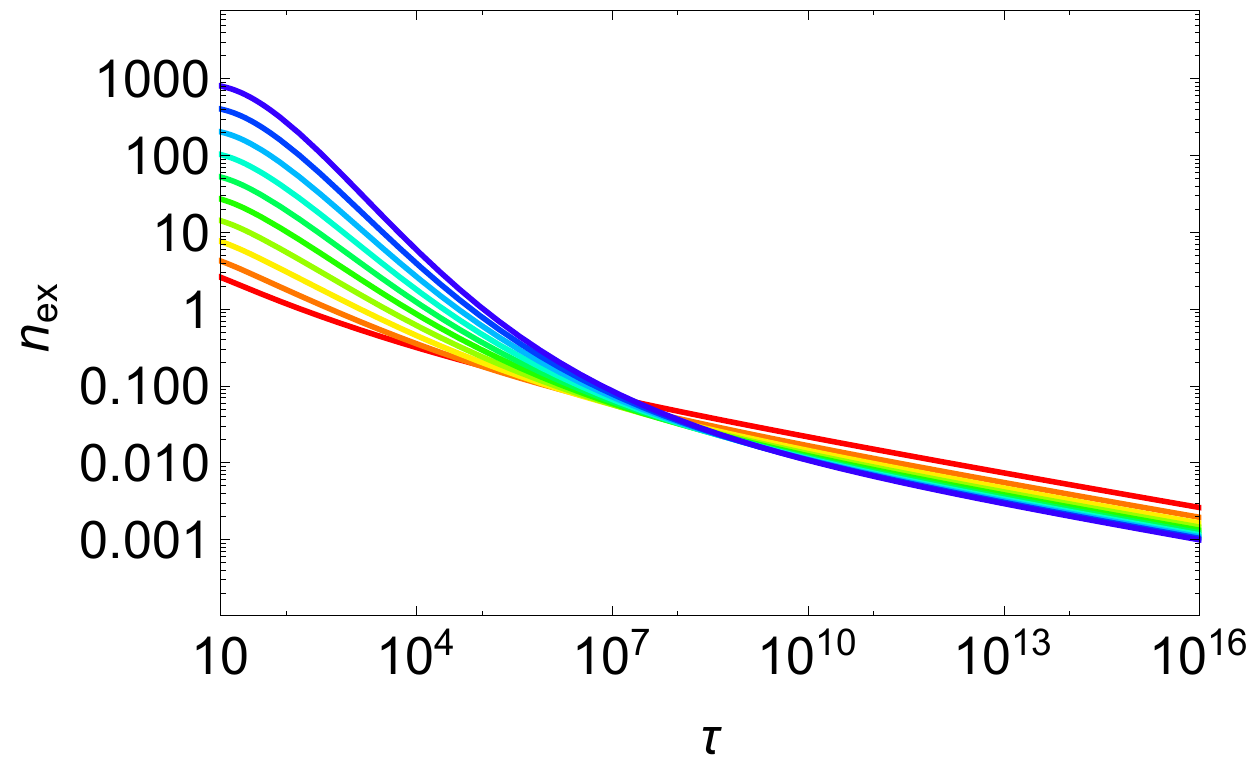}
\caption{Plot of $n_{ex}$ (up to an overall prefactor) given by Eq.~(\ref{nexu}), as a function of $\tau$, for different values of $u$, from $1$ (red line) to $10$ (purple line), increasing by one at a time. For $\tau$ not very large (in the transient regime) the exponent $\gamma(\tau)$ depends on $u$ 
and increases as $u$ increases, 
while for large $\tau$ it becomes universal. We used the values $z=2.69$ and $\nu=2/3$, valid for the Fredkin model in the zero magnetization sector.
}
\label{fig:plot-nex-asy}
\end{figure}
However, as shown in Fig.~\ref{fig:plot-nex-asy}, $n_{ex}$ changes very slowly with $\tau$, and the asymptotic value is reached at very large times $\tau$, where, apart from slowly varying logarithmic corrections, $n_{ex}\sim \tau^{-\gamma}$, with $\gamma$ given by Eq.~(\ref{eq. gamma limit}), explicitly $\gamma=\nu/(2-\nu+2\nu z)$. 
In the transient regime,  
$n_{ex}$ locally scales approximately as $n_{ex}\sim \tau^{-\gamma(\tau)}$, where $\gamma(\tau) = -\frac{\tau \partial_\tau n_{ex}}{n_{ex}}$, which depends also on $u$. 
Finally, we consider also $\beta\neq0$. In this case, calculating the limit of Eq.~\eqref{eq. gamma limit}, we get $\gamma=\frac{\gamma_0+2\beta}{2-\gamma_0}$. \\ 
In conclusion, we get a breakdown of the Kibble-Zurek mechanism due to the fact that the overlaps $\bra{E_l}\partial_\lambda\ket{E_0}$ do not scale all with the same critical exponent.
From a heuristic point of view, we have that the typical size of the defects scales as $\ell\sim \lambda^{-\nu_{e}}$, where $\nu_{e}$ is an effective critical exponent, which depends on the critical exponents $a(k)$, then by separating the dynamics in impulsive and adiabatic as in the Kibble-Zurek mechanism, we get $n_{ex} \sim \tau^{-\nu_{e}/(z\nu+1)}$. In particular, in the limit $\tau\to\infty$ we expect $\nu_e = \nu/(2-\gamma_0)$.

It is worth observing that the scaling formula applies to expectation values $\langle O \rangle$ of observables which are diagonal in the energy basis at the final time. To prove it, we note that at the final time we have $\langle O \rangle = \sum_l O_l \abs{c_l}^2$, where $O_l$ are the diagonal elements of the observable, providing that $O_0=0$. In the thermodynamic limit we get an integral over $k$, and if the main contribution comes from $k\sim 0$, and $O(k_l)=O_l$ is a smooth function, we can approximated $O(k)$ with $O(0)$ in the integral, so that we expect $\langle O \rangle\sim n_{ex}$, namely $\langle O \rangle$ and $n_{ex}$ have the same scaling behavior in $\tau$. 

\begin{figure}
[h!]
\centering
\includegraphics[width=0.75\columnwidth]{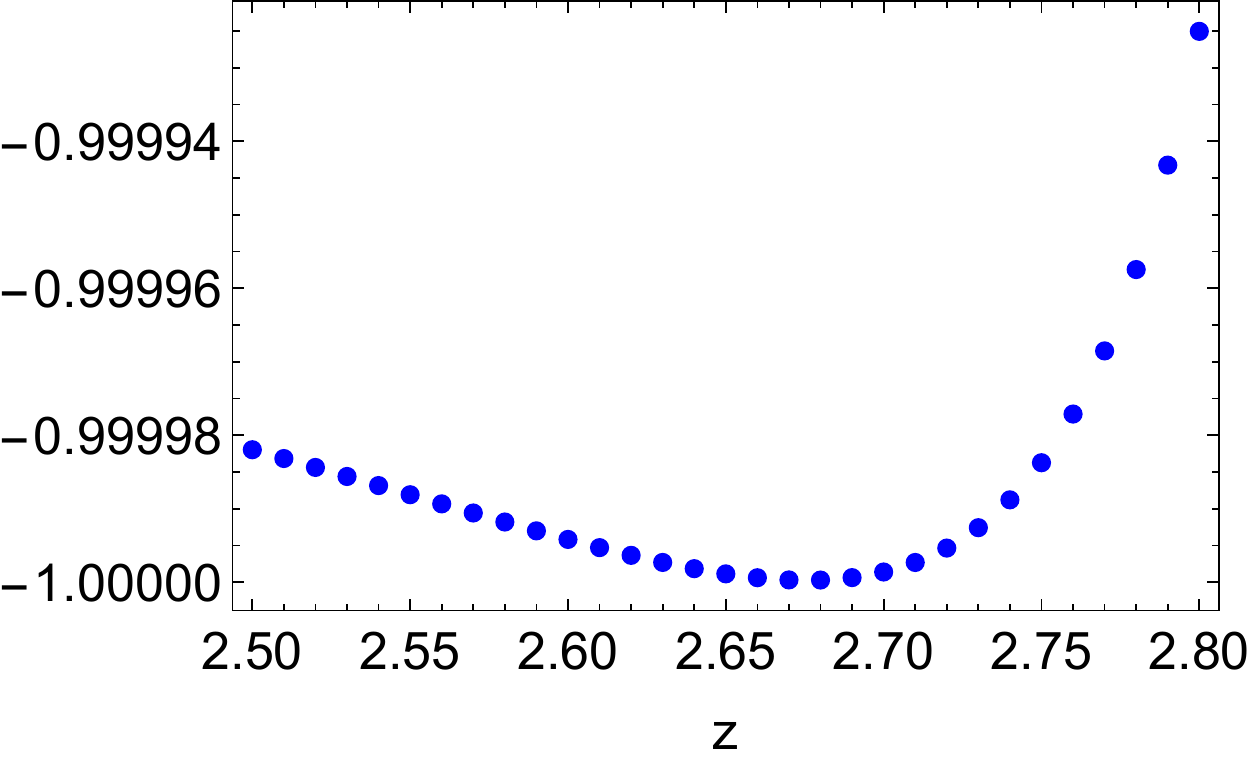}
\caption{Plot of the Pearson correlation coefficient, corresponding to the first excited energy level in the subspace $S_z=0$, between $\ln \Delta_n(z)$ and $\ln n$ for $n\in[5,8]$, by changing $z$ by steps of $0.01$. It is minimum for $z= 2.68$.
}
\label{fig:plot-E1-Sz0}
\end{figure}

\begin{figure}
[h!]
\centering
\includegraphics[width=0.75\columnwidth]{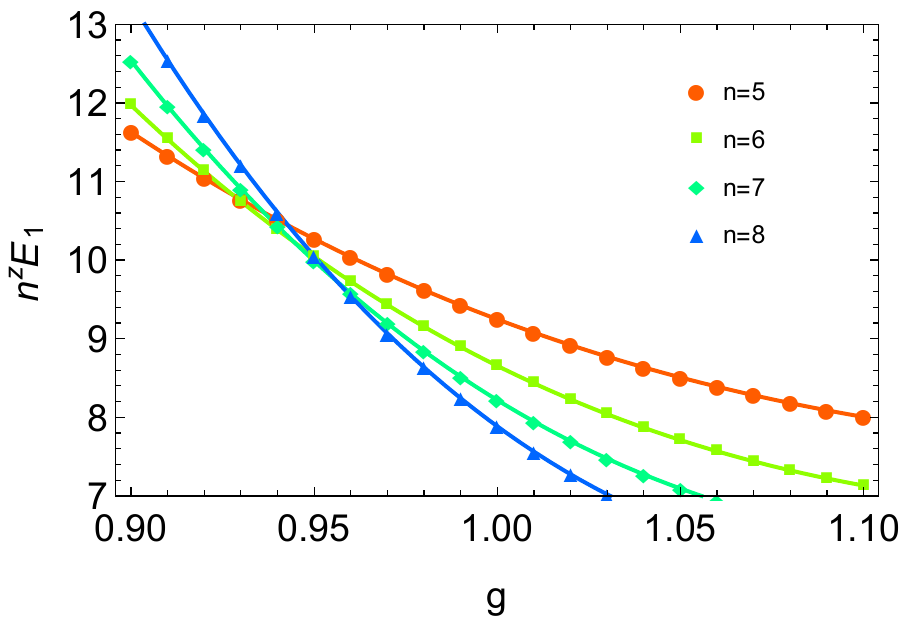}\\
\includegraphics[width=0.75\columnwidth]{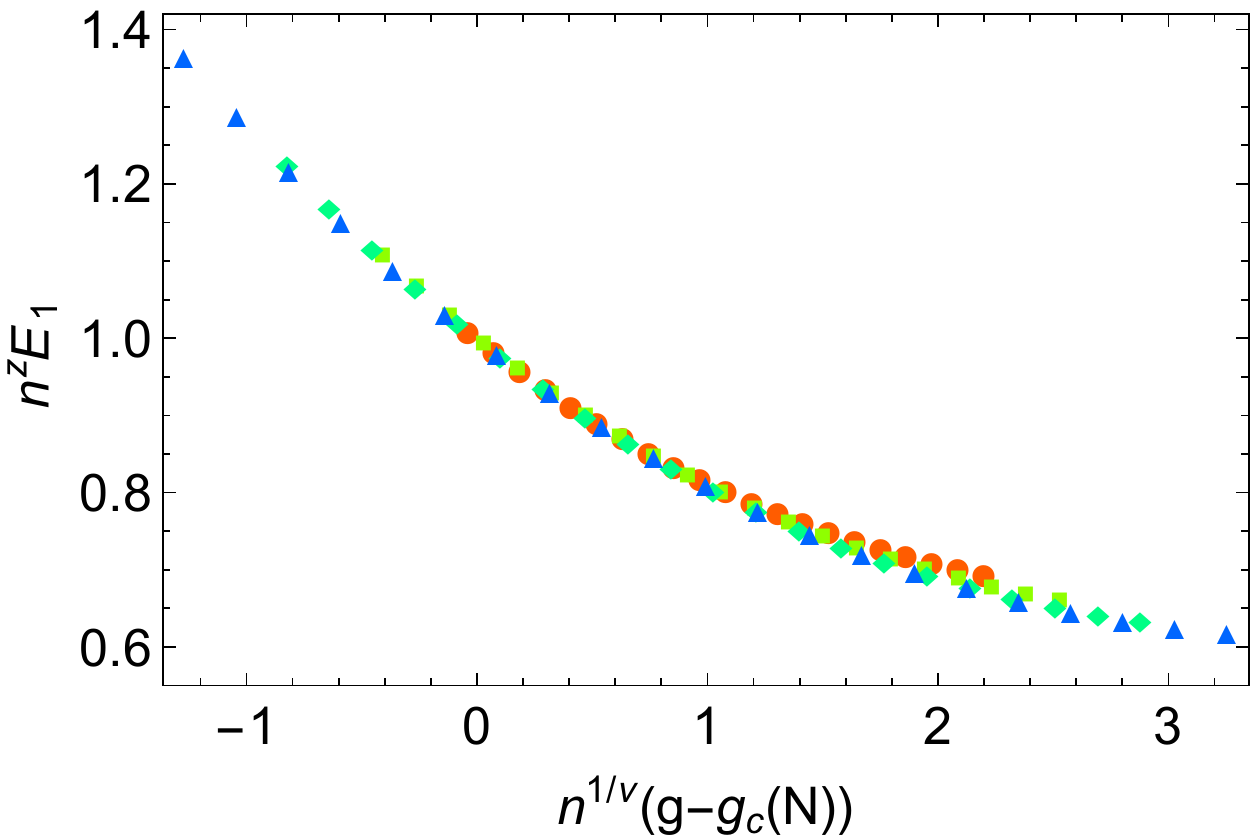}
\caption{ In the top panel, finite size scaling plot of  $n^z E_1$ versus $g$ for different sizes $N=2n$, for $z=2.69$. We note that there is a crossing point at $g_c(N)$, smaller than $g_c=1$. In the bottom panel, finite size scaling plot of $n^z E_1$ versus $ n^{1/\nu}(g-g_c(N))$, for $z=2.69$ and $\nu=2/3$. The values $n^z E_1$ are normalized such that $n^z E_1=1$ for $g=g_c(N)$.  We note that all the points collapse into the same curve.
}
\label{fig:plot-FSS}
\end{figure}

\begin{figure}[h!]
\centering
\includegraphics[width=0.75\columnwidth]{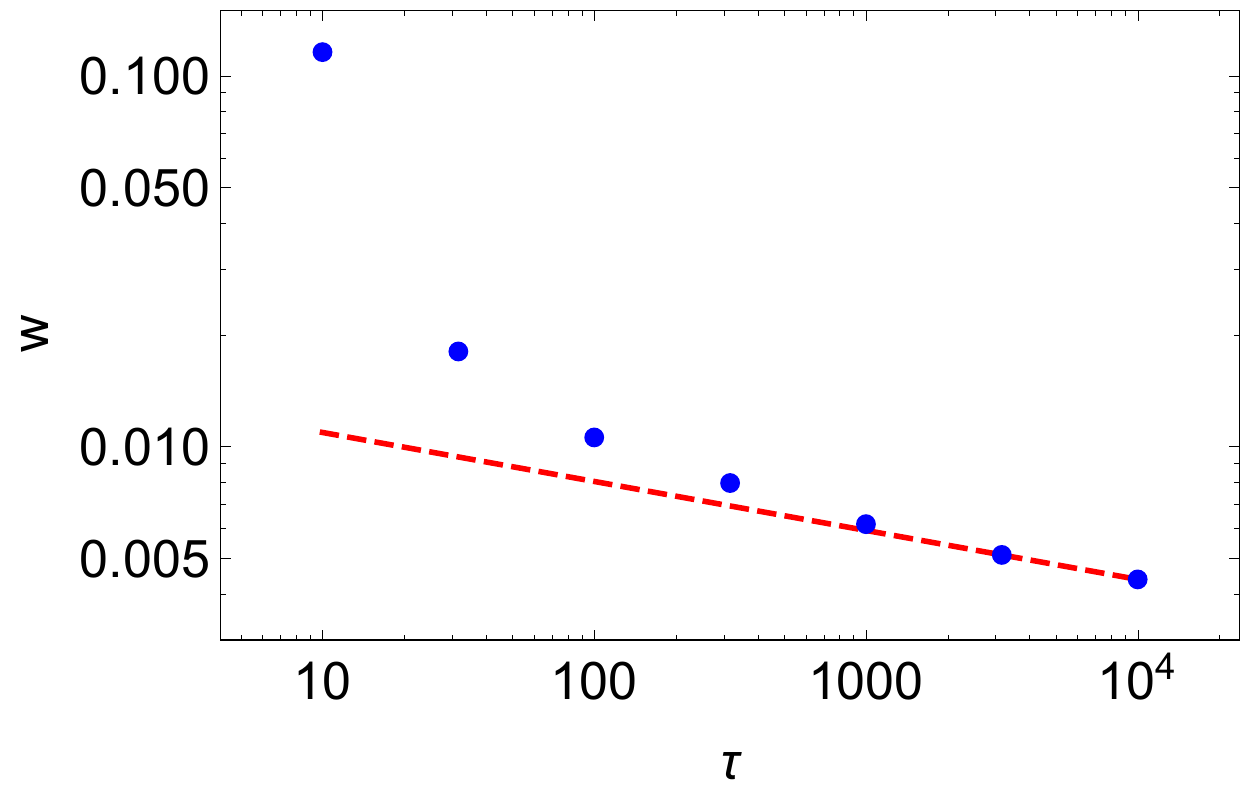} 
\caption{Plot of $w$ as a function of $\tau$, for a Fredkin chain with $N=200$, subjected to an adiabatic quench from $g_{in}=0$ to $g_{fin}=5$. We used matrix product states truncated at the value $D_{max}=10$. The time evolution is calculated by performing a second order Trotter decomposition, with time step $0.1$. The dashed line is a function proportional to $\tau^{-\gamma}$, where $\gamma\approx 0.132$, calculated by fitting the last two data points. This numerical result is in agreement with the theoretical prediction for $\nu=2/3$ and $z=2.69$, which gives $\gamma_0\approx 0.24$ and $\gamma=\gamma_0/(2-\gamma_0)\approx 0.136$.
}
\label{fig:plot-def-MPS}
\end{figure}

\begin{figure}[h!]
\centering
\includegraphics[width=0.75\columnwidth]{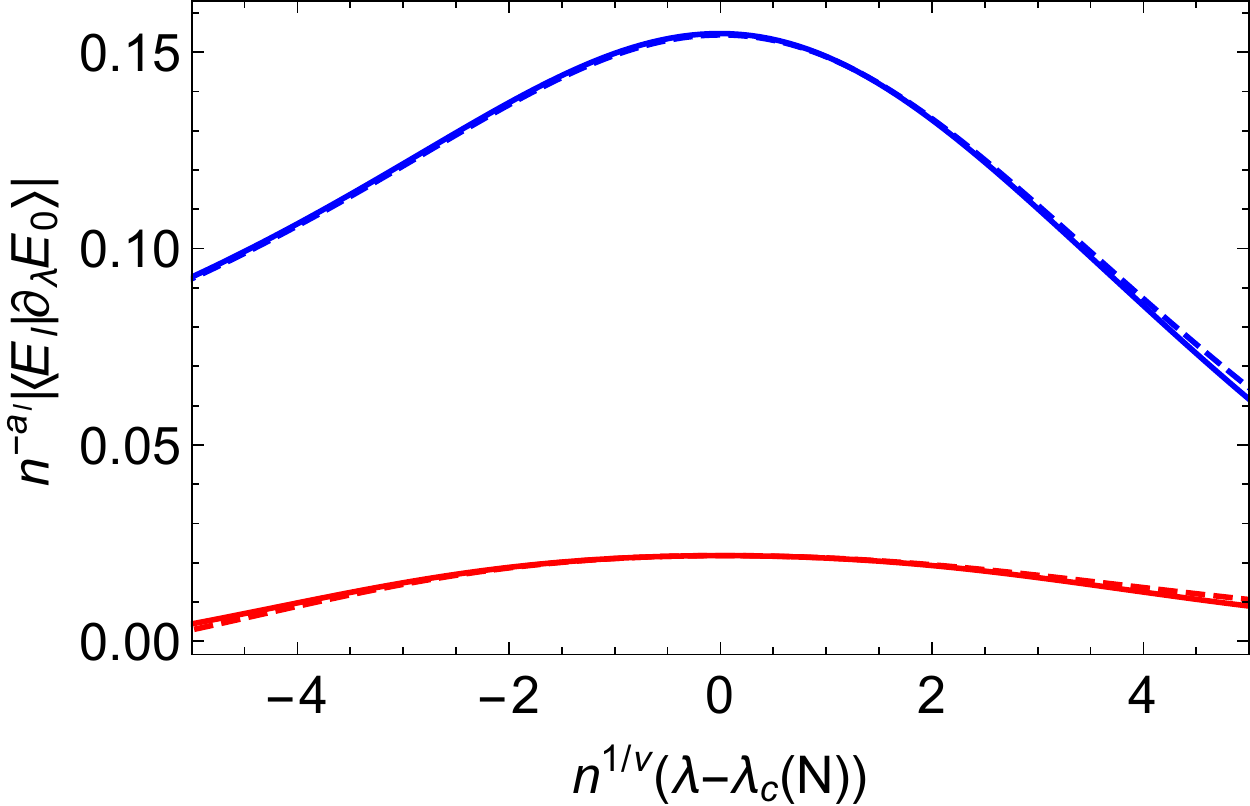}
\caption{Plots of the finite size scaling for the first two overlaps $\bra{E_l}\partial_\lambda \ket{E_0}$. The dashed lines correspond to $n=6$, the solid ones to $n=7$, the blue ones to $l=1$ (first excited level in the zero-spin sector) and the red ones to $l=2$ (second excited level in the zero-spin sector). The value of $\lambda_c(N)$ is calculated as the maximum point of the correspondent overlap. 
We used $a_1=1.65\approx 1/\nu$ and $a_2=2.18$, such that the data points for both $n=6$ and $n=7$ collapse in good approximation in the same curves.
}
\label{fig:plot-fss-over}
\end{figure}

{\it The model.} As a physical example, we consider a chain of $N=2n$ spins $1/2$, described by the Hamiltonian $H(g)= H_\partial + \sum_{j=1}^{N-2} H_j(g)$, where $H_{\partial} = \ket{\downarrow_1}\bra{\downarrow_1}+\ket{\uparrow_N}\bra{\uparrow_N}$ and
\begin{eqnarray}
\nonumber H_j(g) &=& \ket{\uparrow_j} \bra{\uparrow_j}\otimes \ket{s_{j+1,j+2}(g)} \bra{s_{j+1,j+2}(g)} \\
&& + \ket{s_{j,j+1}(g)} \bra{s_{j,j+1}(g)}\otimes \ket{\downarrow_{j+2}} \bra{\downarrow_{j+2}}\,,
\end{eqnarray}
with $\ket{s_{i,j}(g)}=(\ket{\uparrow_i \downarrow_j}- g \ket{\downarrow_i \uparrow_j})/\sqrt{1+g^2}$, where $\ket{\uparrow_j}$ and $\ket{\downarrow_j}$ are eigenstates of $\sigma_j^z$ with eigenvalues $1$ and $-1$, where $\sigma_j^{\alpha}$ with $\alpha=x,y,z$ are the Pauli matrices. As shown in Ref.~\cite{salberger17}, the unique ground-state $\ket{E_0(g)}$ is frustration free and it is a weighted superposition of so-called Dyck paths.
We note that for $g=0$ the ground state is $\ket{\uparrow \downarrow}^{\otimes n}$, conversely,
in the limit $g\to \infty$, the ground level is degenerate and the ground states are of the kind $\otimes_i (\ket{\uparrow}^{\otimes n^\uparrow_i} \otimes \ket{\downarrow}^{\otimes n^\downarrow_i})$, with $n_i^\uparrow\geq 2$ and $n_i^\downarrow\geq 2$, so that $\sum_i n_i^\uparrow = \sum_i n_i^\downarrow = n$, i.e., we get magnetic domains not smaller than two sites.
For the special value $g=1$ we get the Fredkin spin chain~\cite{dellanna16,salberger16}. In this case, the ground-state is a homogeneous superposition of Dyck paths, which displays an entanglement entropy which grows logarithmically with the size of the block. It is shown that for the Fredkin chain, i.e., for $g=1$, the gap closes in the thermodynamic limit as $E_1\sim N^{-z}$, and there are different dynamical critical exponents $z$ depending on the subspace defined by a given value of the magnetization $S_z = \sum_{j=1}^N\sigma^z_j$ (see Ref.~\cite{chen17}). For instance, in the sector with total spin $S_z=0$, by performing a DMRG analysis for large sizes, one gets $z\approx 2.69$ \cite{dellanna16,chen17}. 
Let us derive $z$ by means of an alternative approach which allow us to get a good result already for not too large system sizes. 
Let us consider the ratio $E_1(n)/E_1(n+1)$. For large $n$ we expect that $E_1(n)/E_1(n+1)\sim (1+1/n)^z$. To estimate the value of $z$, we consider the difference $\Delta_n(z) = | E_1(n)/E_1(n+1)-(1+1/n)^z|$. The correct value of $z$ is the one for which $\Delta_n(z)\to 0$ as $n\to \infty$. 
We assume that $\Delta_n(z)\sim n^{-\alpha}$ for the true value of $z$. 
Actually we have numerical evidences that this is the case in the neighbourhood of $z=2.69$. 
By calculating the Pearson correlation coefficient between $\ln \Delta_n(z)$ and $\ln n$ for $n\in[5,8]$, we find that it reaches the value $-1$ for $z\approx 2.68$ (see Fig~\ref{fig:plot-E1-Sz0}). With this approach, therefore, we obtain a value of $z$ very close to the known result, $z=2.69$, already considering small systems sizes.
For $g\sim 1$, in the thermodynamic limit, the gap closes as $E_1 \sim |g-1|^{z \nu}$, so that for $g=g_c=1$ we have a second-order quantum phase transition. Thus, we perform the finite size scaling ansatz~\cite{domb83}
\begin{equation}\label{eq. fss}
E_1(N,g) = N^{-z(N)} f(N^{\frac{1}{\nu}}(g-g_c(N)))
\end{equation}
where $z(\infty)=z$ and $g_c(\infty)=g_c=1$. The pseudo-critical value $g_c(N)$ which is expected to be  $g_c(N)\sim g_c + b N^{-1/\nu}$ for large $N$, can be calculated  as the fixed point of the scale transformation~\cite{domb83,burkhardt85}
\begin{equation}
N^z E_1(N,g) = {N'}^z E_1(N',g')
\end{equation}
thus, by considering $N'=N-2$, $g_c(N)$ is such that $N^z E_1(N,g_c(N)) = (N-2)^z E_1(N-2,g_c(N))$, i.e., corresponds to the value $g$ where the curves $N^z E_1(N,g)$ and $(N-2)^z E_1(N-2,g)$ (as functions of $g$) cross.
The value of the exponent $\nu$ is such that the points $N^z E_1(N,g)$ versus $N^{\frac{1}{\nu}}(g-g_c(N))$ collapse into a single curve $f(v)$ as $N$ goes to infinity. In particular, if the ansatz in Eq.~\eqref{eq. fss} is true, for finite $N$ the points collapse to a curve proportional to $f(v)$. By performing this analysis, we get that  the gap closes with $z\approx2.69$ and $\nu=2/3$ (see Fig.~\ref{fig:plot-FSS}). 
%
After having determined the critical exponents, we proceed by considering a time evolution generated by changing the parameter $g$ linearly in time, as $g(t) =  g_{fin} t/\tau$, in the time interval $t\in [0,\tau]$. The initial state is the ground-state $\ket{E_0(0)} = \ket{\uparrow \downarrow}^{\otimes n}$. 
The time-evolution occurs in the invariant subspace of the Dyck words, having a dimension $C_N = \binom{2n}{n}/(n+1)$ and characterized by a magnetization $S_z$ equal to zero.
Since for $g_{fin}$ very large, in the ground-state there are not magnetic domains smaller than two sites, we consider as defects the domains of only one site. For very large $g$, we get $H_j(g)\sim w_j = \ket{\uparrow \downarrow \uparrow}\bra{\uparrow \downarrow \uparrow}+\ket{\downarrow \uparrow \downarrow}\bra{\downarrow \uparrow \downarrow}$, so the average number of these defects tends to be equal to the average value $\bra{\psi(\tau)}H(g_{fin}) \ket{\psi(\tau)}$, i.e., the irreversible work produced. Based on the observation clarified previously, since the irreversible work scales in the same way as the density of defects, we can consider $w = \sum_j \bra{\psi(\tau)} w_j \ket{\psi(\tau)} /N$ to determine the scaling of defects, which is easier to calculate, resorting to the matrix product state and second order Trotter decomposition~\cite{schollwock11} (see Fig.~\ref{fig:plot-def-MPS}). 
If the conventional theory of the Kibble-Zurek mechanism could be applied, we would expect $w\sim \tau^{-\gamma_0}$, with $\gamma_0 = \nu/(z\nu+1)<1$, if $z>1$. However, as shown in Fig.~\ref{fig:plot-def-MPS}, we find that, for not very large $\tau$, in the transient regime, $w$ 
scales with a non-universal exponent $\gamma$ larger than one. 
Conversely, $w\sim \tau^{-\gamma}$ scales with an exponent $\gamma$ (apart from a slow varying logarithmic function) which approaches the universal value $\gamma=\gamma_0/(2-\gamma_0)$ for large $\tau$, as shown in  Fig.~\ref{fig:plot-def-MPS}.
This anomalous dynamical behavior for the deformed Fredkin model can be explained requiring that the overlaps $\bra{E_l}\partial_\lambda \ket{E_0}$ scale with level-dependent critical exponents $a(k_l)\sim 1/\nu + u k_l$, as shown in Fig.~\ref{fig:plot-fss-over}. 

%

{\it Conclusions.} The characterization of quantum phase transitions is of particular importance in condensed matter physics. In particular, the Kibble-Zurek mechanism has been thoroughly investigated over the years. Here, we considered the possibility that one of the two scaling formulas required to get the universal dynamics of the Kibble-Zurek mechanism can be generalized, so that we can have a violation of this mechanism. Nevertheless, we have shown that also in this case, we can get a universal dynamics, although different from the Kibble-Zurek one. We provided an example where this anomalous universal behavior can occur. 
We showed that the Fredkin spin chain has the required features. By performing an explicit numerical calculation for the irreversible work, related to the density of defects, we have found a scaling law in perfect agreement with what was expected by our theory, namely, a universal adiabatic dynamics different from the Kibble-Zurek prediction. 

Recently, several exceptions to the Kibble-Zurek mechanism have been found, e.g., in the presence of long-range interactions \cite{defenu,noi}, or  
across a localization-delocalization transition \cite{igloi}, and, in general, an explanation of an anti-Kibble-Zurek behavior \cite{dutta} and a different scaling in fast quenches \cite{zeng} have also been investigated. 
{\color {black}The origin of the Kibble-Zurek violation, reported in our work, relies on the properties of the spectrum, 
however, how to detect this peculiar behavior from the Hamiltonian is still an open question.}
We hope that our work can give a further boost to current research in this direction, 
and can be useful for a deeper understanding of a class of still unexplored critical systems, 
inspiring further investigations and applications in the fields of quantum phase transitions and out-of-equilibrium phenomena.

{\it Acknowledgements.} The authors acknowledge financial support from the project BIRD 2021 "Correlations, dynamics and topology in long-range quantum systems" of the Department of Physics and Astronomy, University of Padova, and from the European Union - Next Generation EU within the "National Center for HPC, Big Data and Quantum Computing" (Project No. CN00000013, CN1 Spoke 10 Quantum Computing).


\begin{thebibliography}{99}
\bibitem{sachdevbook} S. Sachdev, "Quantum Phase Transition" (Cambridge University Press, Cambridge, 1999).
\bibitem{zurek05} W. H. Zurek, U. Dorner and P. Zoller, 
"Dynamics of a Quantum Phase Transition", 
Phys. Rev. Lett. 95, 105701 (2005).
\bibitem{dziarmaga05} J. Dziarmaga, 
"Dynamics of a Quantum Phase Transition: Exact Solution of the Quantum Ising Model", 
Phys. Rev. Lett. 95, 245701 (2005).
\bibitem{polkovnikov05} A. Polkovnikov, 
"Universal adiabatic dynamics in the vicinity of a quantum critical point", 
Phys. Rev. B 72, 161201(R) (2005).
\bibitem{kibble76} T. W. B. Kibble, 
"Topology of cosmic domains and strings", 
J. Phys., A9, 1387 (1976).
\bibitem{zurek85} W. H. Zurek, 
"Cosmological experiments in superfluid helium?", 
Nature 317, 505 (1985).
\bibitem{polkovnikov11} A. Polkovnikov, K. Sengupta, A. Silva and M. Vengalattore, 
"Colloquium: Nonequilibrium dynamics of closed interacting quantum systems", 
Rev. Mod. Phys. 83, 863 (2011).
\bibitem{delcampo14} A. del Campo and W. H. Zurek, 
"Universality of phase transition dynamics: Topological defects from symmetry breaking", 
Int. J. Mod. Phys. A 29, 1430018 (2014).
\bibitem{salberger17} O. Salberger, T. Udagawa, Z. Zhang, H. Katsura, I. Klich and V. Korepin, 
"Deformed Fredkin Spin Chain with Extensive Entanglement", 
J. Stat. Mech. (2017) 063103.
\bibitem{pachos} J.K. Pachos and E. Rico, "Effective three-body interactions in triangular optical lattices", Phys. Rev. A 70, 053620 (2004).
\bibitem{dellanna16} L. Dell'Anna, O. Salberger, L. Barbiero, A. Trombettoni and V. E. Korepin, 
"Violation of cluster decomposition and absence of light cones in local integer and half-integer spin chains", 
Phys. Rev. B 94, 155140 (2016).
\bibitem{salberger16} O. Salberger and V. Korepin, 
"Entangled spin chain", 
Rev. Math. Phys. 29, 1750031 (2017).
\bibitem{domb83} C. Domb, M. Green, and J. Lebowitz, "Phase Transitions and Critical Phenomena", Vol. 8 (Academic Press, New York, 1983).
\bibitem{chen17} X. Chen, E. Fradkin and W. Witczak-Krempa, 
"Gapless quantum spin chains: multiple dynamics and conformal wavefunctions", 
J. Phys. A: Math. Theor. 50 464002 (2017).
\bibitem{burkhardt85} T. W. Burkhardt and I. Guim, 
"Finite-size scaling of the quantum Ising chain with periodic, free, and antiperiodic boundary conditions", 
J. Phys. A: Math. Gen. 18, L33 (1985).
\bibitem{schollwock11} U. Schollw\"{o}ck, 
"The density-matrix renormalization group in the age of matrix product states", 
Ann. Phys. 326, 96 (1) (2011).
\bibitem{defenu} N. Defenu, G. Morigi, L. Dell'Anna, and T. Enss, 
"Universal dynamical scaling of long-range topological superconductors", 
Phys. Rev. B 100, 184306 (2019).
\bibitem{noi} G. Francica, L. Dell'Anna, "Correlations, long-range entanglement and dynamics in long-range Kitaev chains", Phys. Rev. B 106, 155126 (2022).
\bibitem{igloi} G. Ro\'osz, U. Divakaran, H. Rieger, F. Igl\'oi, 
"Nonequilibrium quantum relaxation across a localization-delocalization transition", 
Phys. Rev. B 90, 184202 (2014).
\bibitem{dutta} A. Dutta, A. Rahmani, and A. del Campo, 
"Anti-Kibble-Zurek Behavior in Crossing the Quantum Critical Point of a Thermally Isolated System Driven by a Noisy Control Field", 
Phys. Rev. Lett. 117, 080402 (2016).
\bibitem{zeng} H.-B.  Zeng, C.-Y. Xia, and A. del  Campo, 
"Universal breakdown of Kibble-Zurek scaling in fast quenches across a phase transition", 
arXiv:2204.13529 (2022).
\end{thebibliography}
\end{document}